\documentclass[prd, superscriptaddress,nofootinbib,twocolumn]{revtex4}
\usepackage{graphicx,natbib}
\usepackage{url}
\usepackage{color}
\usepackage{rotating}
\usepackage{amssymb,amsmath}

\newcommand{\Mobs}{M_{\rm obs}}
\newcommand{\Mbias}{M^{\rm bias}}
\newcommand{\Mth}{M^{\rm th}}
\newcommand{\siglnM}{\sigma_{\ln M}}

\newcommand{\sigz}{\sigma_{z}}
\newcommand{\Msun}{M_{\odot}}

\newcommand{\FoM}{{\rm FoM}}
\newcommand{\gFoM}{{\gamma \rm FoM}}
\begin{document}

\title{Constraining Dark Energy with Clusters: Complementarity with Other Probes}

\author{Carlos Cunha}
\affiliation{Department of Physics, University of Michigan, 
450 Church St, Ann Arbor, MI 48109-1040}

\author{Dragan Huterer}
\affiliation{Department of Physics, University of Michigan, 
450 Church St, Ann Arbor, MI 48109-1040}

\author{Joshua A.\ Frieman}
\affiliation{Center for Particle Astrophysics, Fermi National Accelerator
  Laboratory, P. O. Box 500, Batavia, IL 60510; \\
Kavli Institute for Cosmological Physics, The University of Chicago, 5640 S. Ellis Ave., 
Chicago, IL 60637}

\date{\today}

\begin{abstract}
The Figure of Merit Science Working Group (FoMSWG) recently forecast the 
constraints on dark energy that will be achieved prior to the Joint Dark 
Energy Mission (JDEM) by ground-based experiments that exploit
baryon acoustic oscillations, type Ia supernovae, and weak 
gravitational lensing. 
We show that cluster counts from on-going and near-future surveys should 
provide robust, complementary dark energy constraints. 
In particular, we find that optimally combined optical and Sunyaev-Zel'dovich 
effect cluster surveys should improve the Dark Energy Task Force (DETF) 
figure of merit for pre-JDEM projects by a factor of two even without prior 
knowledge of the nuisance parameters in the cluster mass-observable relation.
Comparable improvements are achieved in the forecast precision of parameters 
specifying the principal component description of the dark energy equation 
of state parameter as well as in the growth index $\gamma$. 
These results indicate that cluster counts can play an important complementary 
role in constraining dark energy and modified gravity even if the associated 
systematic errors are not strongly controlled.
 \end{abstract}

\maketitle
\section{Introduction}\label{sec:intro}

Counts of galaxy clusters are a potentially very powerful technique to probe
dark energy and the accelerating universe (e.g.\ \cite{FTH,sah09,mar06,voi05,pie03,bat03,ros02,hai01,hol01}).  The idea
is an old one: count clusters as a function of redshift (and, potentially,
mass), and compare to theoretical prediction which can be obtained either
analytically or numerically.  Recently, Rozo et al.\ \cite{Rozo09} have
obtained very interesting constraints on $\sigma_8$ from Sloan Digital Sky
Survey (SDSS) cluster samples using the relation between mass and optical
richness (the number of red-sequence galaxies in the cluster above a
luminosity threshold).  This follows recent dark energy constraints using
optical \cite{gla07} and X-ray observations of clusters \cite{Mantz07,Vikhlinin08,hen09}.

In this paper we calculate the potential of cluster counts to improve combined
constraints from the other three major probes of dark energy: baryon acoustic
oscillations (BAO), type Ia supernovae (SNIa) and weak gravitational lensing
(WL). We are motivated by the recently released report of the Figure of Merit
Science Working Group (FoMSWG; \cite{FoMSWG}) that studied and recommended
parametrizations and statistics best suited to addressing the power of
cosmological probes to measure properties of dark energy. While the FoMSWG
report was mainly aimed at figures of merit to be used in the upcoming
competition for the Joint Dark Energy Mission (JDEM) space telescope, the
applicability of its results and recommendations is general. 

We address quantitatively how ongoing and upcoming cluster
surveys, in particular the South Pole Telescope (SPT, \cite{ruh04}) and the
Dark Energy Survey (DES, \cite{des05}), can strengthen the combined
``pre-JDEM'' constraints on dark energy considered in the FoMSWG report ---
that is, combined constraints expected around the year 2016. To model cluster
counts, we utilize recent results from Cunha \cite{cun08} which optimally
combine future optical and Sunyaev-Zel'dovich (SZ) observations of clusters to
estimate the constraints on dark energy.

%%%%%%%%%%%%%%%%%%%%%%%%%%%%%%%%%%%%%%%%%%%%%%%%%%%%%%%%%%%%%%%%%%%%%%%%%%%%%%%%%%%%
\section{Information from cluster counts and clustering}\label{sec:counts}

The subject of deriving cosmological constraints from cluster number counts
and clustering of clusters has been treated extensively in the literature (see
e.g.\ \cite{cun08,lim04,lim05,lim07, wu08}).  In this work we use
cross-calibration for two observable proxies for mass: Sunyaev-Zel'dovich flux
(henceforth SZ), and optical observations --- which identify clusters via
their galaxy members --- (henceforth OPT).  
While we focus on optical and SZ surveys, our results
are applicable to combinations of any cluster detection techniques.  In
particular, planned X-ray surveys such as eRosita \cite{pre07}, WFXT
\cite{gia09}, and IXO \cite{vik09} will have mass sensitivity competitive
with, and complementary to, the SZ and optical surveys.%}

Our approach closely follows that in \cite{cun08} and we refer the reader to
that publication for basic details.  In brief, cluster counts in a bin of the
observables are calculated by integrating the mass function $dn/dM$ over mass,
volume, and the observable proxy in the appropriate range. 
We adopt the Jenkins mass function in this work, though results are weakly dependent
on this choice.  
Clustering is given by the sample covariance of the mean counts in different redshift bins.
The contribution of clustering to the constraints is very small when cross-calibration
is used \cite{cun08}.
We allow for scatter in both the relation between mass and the observable
proxy, and the relation between true and estimated photometric redshifts.
Results from both simulations (e.g.\ \cite{sha07,kra06}) and observations
(e.g. \citep{ryk08b,evr08,ryk08a}) suggest that the mass-observable relations
can be parametrized in simple forms with lognormal scatter of the
mass-observable about the mean relation.  Other works (see e.g.\ \cite{coh07})
suggest that the distribution of galaxies in halos may be more complicated.
We assume lognormal scatter for the mass-observable relation as well as for
the photometric redshift errors.
We have neglected any theoretical uncertainties in the mass function, 
galaxy bias, and photometric redshifts,  all of which must be independently 
known to a few percent so as not to  affect cosmological constraints \cite{cun09, lim07}.

We fix the photo-z scatter to $\sigz = 0.02$, the expected overall scatter of
cluster photo-z's in the Dark Energy Survey \citep{des05}.  Our ``theorist's
observable'' quantity, which we feed into the Fisher matrix formalism to
obtain constraints on cosmological parameters, is the covariance of the counts
--- defined as the sample covariance plus the shot noise variance --- in different
redshift bins.

We adopt the same surveys and parametrizations described in \cite{cun08},
namely, an SZ and an OPT survey on the same 4000 sq.\ deg.\ patch of sky.

 Let $\Mobs$ be the observable proxy for mass (from either SZ or OPT survey).
For the SZ survey, we define the $\Mobs$ threshold for detection to be
 $\Mth=10^{14.2}h^{-1}\Msun$, complete up to $z=2$, based on the
projected sensitivity of the South Pole Telescope.   We parametrize the mass bias
(the difference between the true mass and SZ $\Mobs$), and the variance in the
mass-observable relation, respectively as
\begin{eqnarray}
  {\rm ln}\Mbias(z)&=&{\rm ln}\Mbias_0 + a_1\ln(1+z) \label{eqn:mbiasdefsz}\\[0.2cm]
\siglnM^2(z)&=&\sigma_{0}^2 + \sum_{i=1}^{3}b_iz^i
\label{eqn:msigdefsz}
\end{eqnarray}
Fiducial mass values of all nuisance parameters are zero, except for the
scatter which is set to $\sigma_{0}=0.25$ in the fiducial model.  Our choice
of scatter is somewhat conservative given recent studies which suggest that
$\sigma_{0}<0.2$ (see e.g. \cite{mel06}).  However, hydrodynamics simulations by
\cite{hal07} find a scatter of $(+32\%, -16\%)$ about the median for clusters
of $M\sim3.0\times 10^{14} \Msun$, which matches our choice.  In total, there
are six nuisance parameters for the mass bias and scatter (${\rm ln}\Mbias_0$,
$a_1$, $\sigma_{0}^2$, $b_i$).

\begin{figure}[!t]
\includegraphics[scale=0.30]{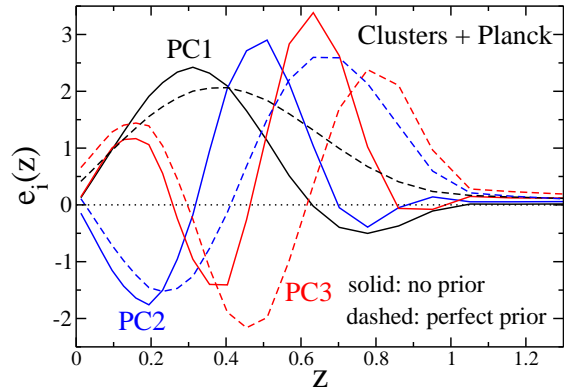}
\vspace{-0.5cm}
\caption{First three principal components for the OPT+SZ cluster
  survey with Planck priors. Solid lines refer
  to the case with no prior on the 16 nuisance parameters, while the dashed
  lines correspond to the case of perfectly known nuisance parameters.}
\label{fig:PC_clus}
\end{figure}

For the optical survey the mass threshold of the observable is set to
$\Mth=10^{13.5}h^{-1}\Msun$ and the redshift limit is $z=1$, corresponding to
the projected sensitivity of the Dark Energy Survey.  Different studies
suggest a wide range of scatter for optical observables, ranging from a
constant $\siglnM=0.5$ \citep{wu08} to a mass-dependent scatter in the range
$0.75 < \siglnM < 1.2$ \citep{bec07}.  Using weak lensing and X-ray analysis
of MaxBCG selected optical clusters, Ref.~\cite{roz08a} estimated a lognormal
scatter of $\sim 0.45$ for $P(M|\Mobs)$, where $M$ was determined using weak
lensing and $\Mobs$ was an optical richness estimate.
We choose a fiducial mass scatter of $\siglnM=0.5$ and allow for a cubic
evolution in redshift and mass:
\begin{eqnarray}
{\rm ln}\Mbias(\Mobs,z)&=&{\rm ln}\Mbias_0 + a_1\ln(1+z)\nonumber \\[0.2cm]
&+&a_2({\rm ln}\Mobs-{\rm ln}M_{\rm pivot}) \label{eqn:mbiasdef}\\[0.0cm]
\siglnM^2(\Mobs,z)&=&\sigma_{0}^2 + \sum_{i=1}^{3}b_iz^i \nonumber \\[-0.2cm]
&+& \sum_{i=1}^{3}c_i({\rm ln}\Mobs-{\rm ln}M_{\rm pivot})^i \label{eqn:msigdef}
\end{eqnarray}
\noindent We set $M_{\rm pivot}=10^{15}h^{-1} \Msun$. 
In all, we have 10 nuisance parameters for the optical mass errors 
(${\rm ln}\Mbias_0$, $a_1$, $a_2$, $\sigma_{0}^2$, $b_i$, $c_i$).

There are few, if any, constraints on the number of parameters necessary to
realistically describe the evolution of the variance and bias with mass.
Ref.\ \cite{lim05} shows that a cubic evolution of the mass-scatter with
redshift captures most of the residual uncertainty when the redshift evolution
is completely free (as assumed in the Dark Energy Task Force (DETF) report
\cite{DETF}).  Note too that we employ more nuisance parameters to describe
the optical survey than the SZ survey because the former is expected to have a
more complicated selection function.  For the cross-calibration analysis, we
assume the correlation coefficient between optical and SZ scatter $\rho$,
defined in \cite{cun08}, to be fixed to zero; the same paper shows that the
cross-calibration results are insensitive to the value of $\rho$ for $\rho \in
[-1, 0.6]$.

In total, we use 6+10=16 nuisance parameters to describe the systematics of
the combined OPT+SZ cluster survey.  While generous, this
parametrization assumes a lognormal distribution of the mass-observable relation 
that may fail for low-masses. 
We have also implicitly assumed that selection effects can be described by the
bias and scatter of the mass-observable relation.  By the year 2016, we expect
significant progress in simulations of cluster surveys that will allow us to
better parametrize the cluster selection errors.

\begin{figure*}[!t]
\includegraphics[scale=0.30]{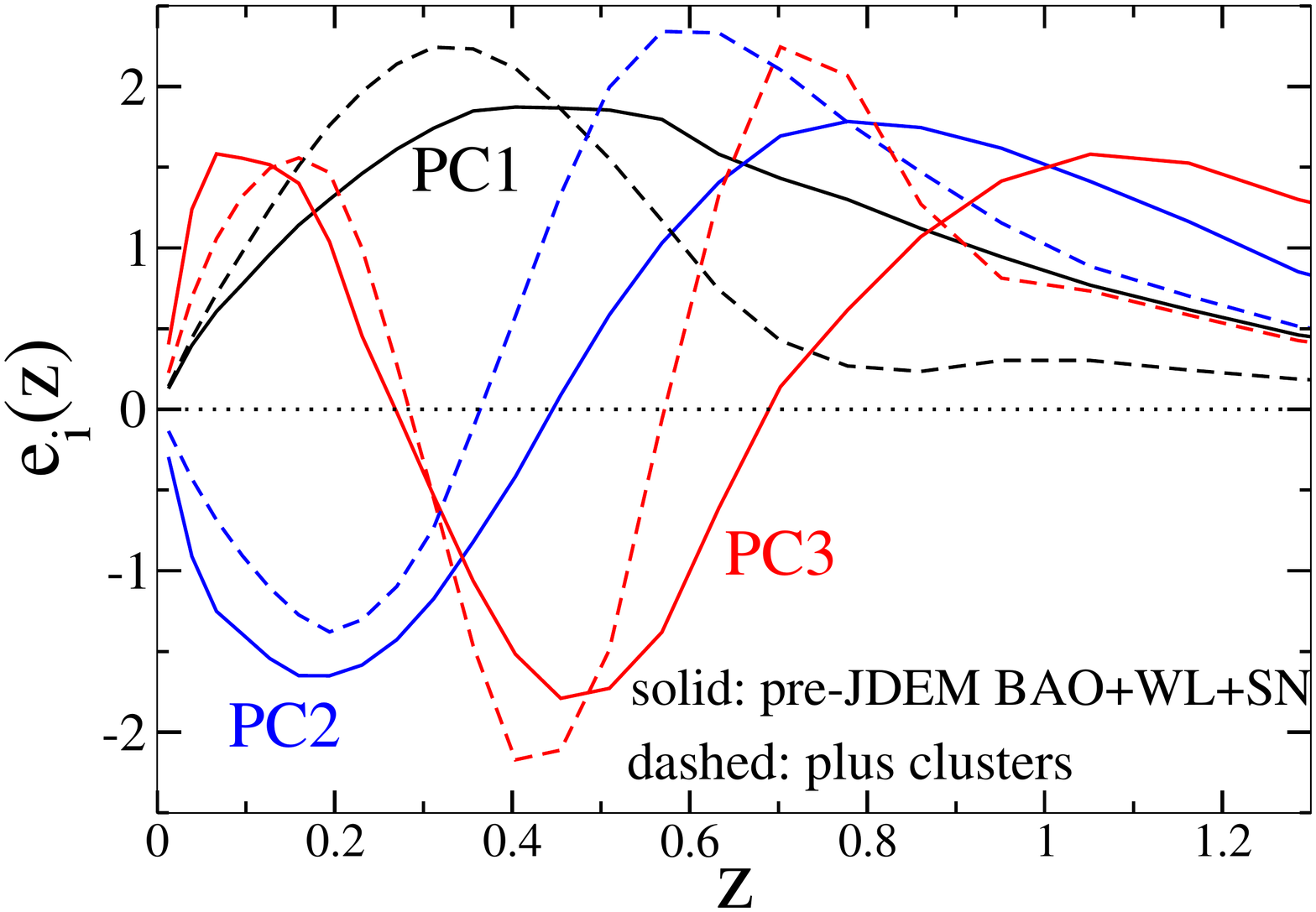}
\includegraphics[scale=0.30]{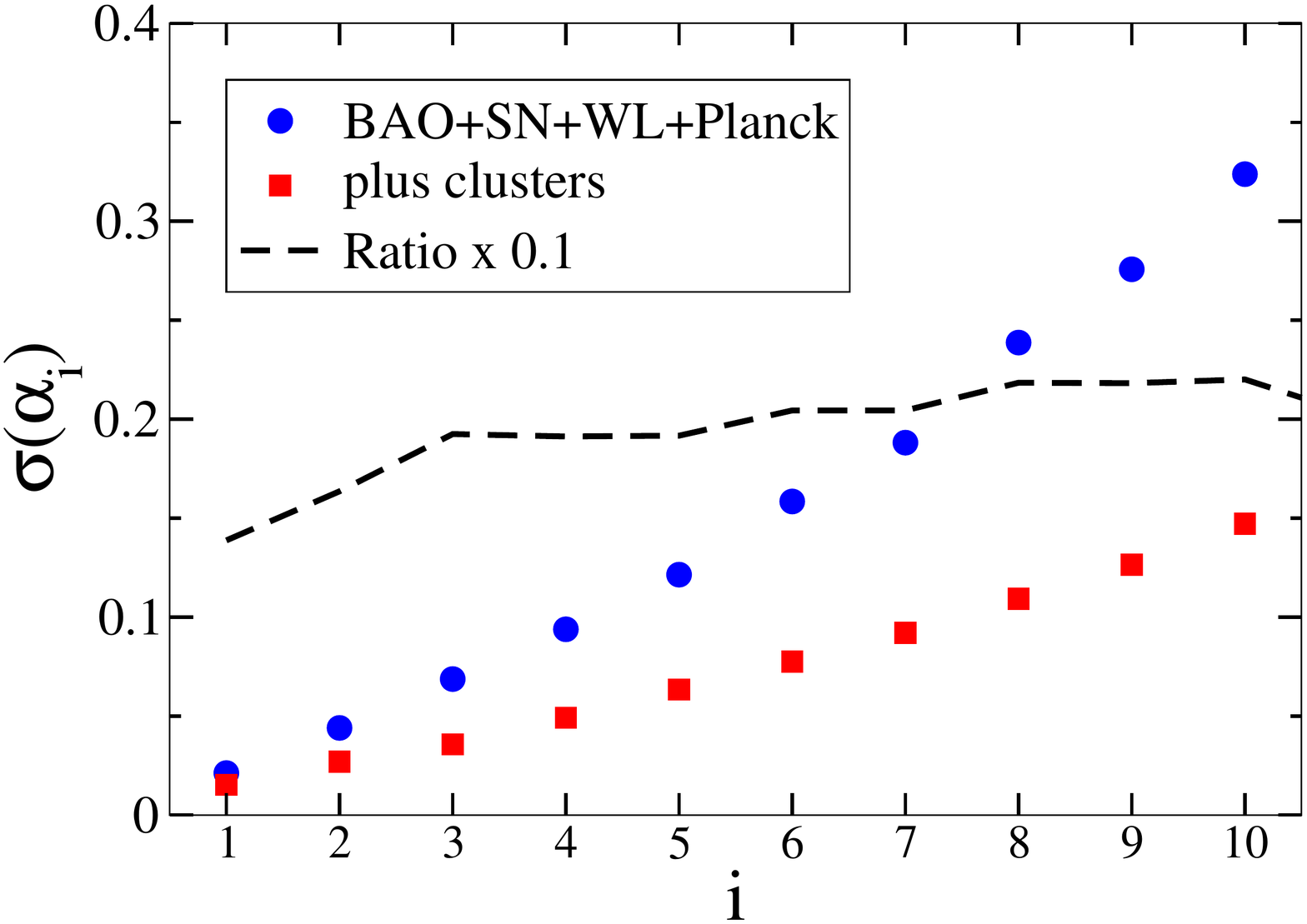}
\vspace{-0.5cm}
\caption{Left panel: First three principal components for the
  SNIa+BAO+WL+Planck pre-JDEM combination alone (solid curves) and for the same combination
  with the addition of clusters (dashed curves). For the latter we assume OPT+SZ cluster
  survey with flat (i.e.\ uninformative) external priors on nuisance
  parameters.  Right panel: Uncertainty in eigencoefficients of $w(a)$,
  $\sigma(\alpha_i)$, for the SNIa+BAO+WL+Planck pre-JDEM combination alone (circular points),
  and for the same combination with the addition of clusters (square points). We also show the
  ratio of the improvement in each eigencoefficient when clusters are added
  (dashed line - scaled down by a factor of 10 for clarity). }
\label{fig:PC_combined}
\end{figure*}

%%%%%%%%%%%%%%%%%%%%%%%%%%%%%%%%%%%%%%%%%%%%%%%%%%%%%%%%%%%%%%%%%%%%%%%%%%%%%%%%%%%%
\section{Complementary probes and Figures of Merit}\label{sec:jdemprobes}

To model the power of complementary probes of dark energy, we adopt the
pre-JDEM information (that is, combined information projected around year
2016) based on estimates of the Figure of Merit Science Working Group
\cite{FoMSWG}. These estimates include information from BAO, SNIa, WL and the
Planck CMB satellite. We use these probes in combination, without or with
clusters.  Note that 
systematic errors have been included in all of these methods
(see Ref.~\cite{FoMSWG}).

The FoMSWG figures of merit are described in the FoMSWG paper \cite{FoMSWG}
and we review them here very briefly.  There are a total of 45 cosmological
parameters, 36 of which describe the equation of state $w(z)$ while the others
are mostly standard cosmological parameters (plus a couple of nuisance ones
that have not been explicitly marginalized over).  One figure-of-merit is the
area in the $w_0$-$w_a$ plane \cite{Huterer_Turner,DETF}, where
$w(z)=w_0+w_a(1-a)= w_p+w_a(a_p-a)$ and where $w_p$ and $a_p$ are the ``pivot''
parameter and the scale factor; we adopt $\FoM\equiv 1/(\sigma(w_p)\times
\sigma(w_a))$. The growth of density perturbations is described by a single
parameter, the growth index $\gamma$, which is a free parameter in the fitting
function for the linear growth of perturbations \cite{Linder_growth}. The
figure-of-merit in the growth index is simply its inverse marginalized error,
$\gFoM\equiv 1/\sigma(\gamma)$.

A much richer (and less prone to biases) description of the equation
of state is achieved through computing the principal components (PCs) of dark
energy \cite{Huterer_Starkman}, $e_i(z)$
\vspace{-0.2cm}
\begin{equation}
1+w(a) = \sum_{i=0}^{35} \alpha_i e_i(a),
\label{eq:PC_expansion}
\end{equation}
where $\alpha_i$ are coefficients, and $e_i(a)$ are the eigenvectors (see
\cite{FoMSWG} for details).  The associated figure of merit consists of
presenting the shapes $e_i(z)$ in redshift and computing the associated
accuracies $\sigma(\alpha_i)$ with which the coefficients can be measured
\cite{FoMSWG}.

Combining the different cosmological probes is achieved by adding their
associated Fisher matrices.  We add the $45\times 45$ Fisher matrix for
clusters (marginalized over the mass nuisance parameters) to the combined BAO+SNIa+WL+Planck 
Fisher matrix and report the improvement in the figures of merit and
accuracies in the PCs as well as shapes of the new PCs. 

%%%%%%%%%%%%%%%%%%%%%%%%%%%%%%%%%%%%%%%%%%%%%%%%%%%%%%%%%%%%%%%%%%%%%%%%%%%%%%%%%%%%
\section{Results}\label{sec:res}
 
\begin{figure*}[!t]
\includegraphics[scale=0.30]{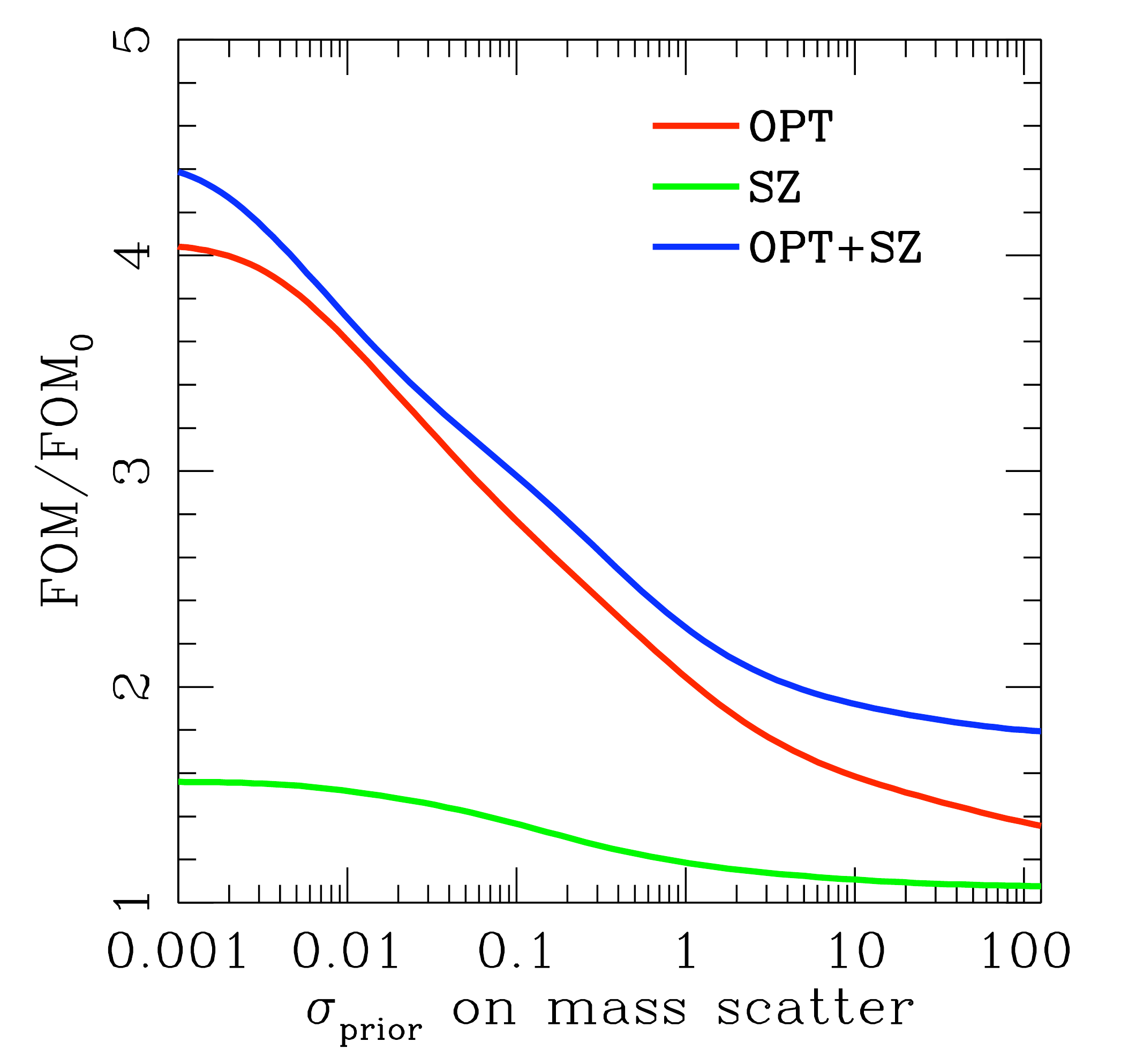}
\includegraphics[scale=0.30]{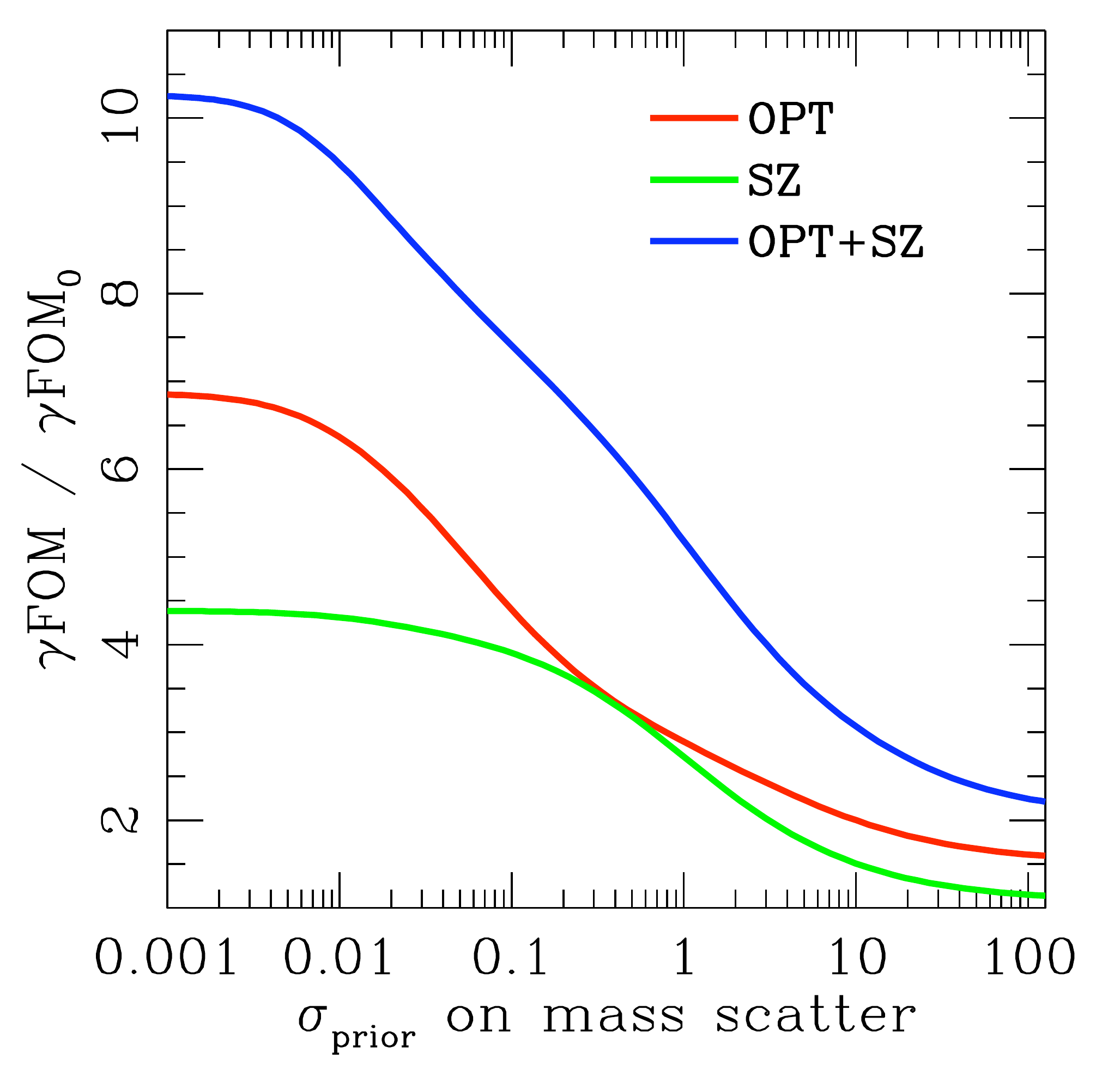}
\vspace{-0.3cm}
\caption{Improvement in the figures of merit relative to the pre-JDEM
  combination (BAO+SNIa+WL+Planck) when cluster information is added.  In both
  panels we add a uniformly increasing prior on the nuisance parameters as
  explained in the text, and show (on the x-axis) the
    effective resulting prior on the scatter in mass $\siglnM(M, z)$, that is,
    the (square root of the) left-hand sides of Eqs.~(\ref{eqn:msigdefsz}) or
    (\ref{eqn:msigdef}) that correspond to the prior values of the nuisance
    parameters on the right-hand sides at $z=1.0$ and $M=10^{15}h^{-1} \Msun$.
    For the OPT+SZ case, the uncertainty in the optical scatter is shown on
    the x-axis.  The left panel shows the improvements in the (DETF) $\FoM$,
  while the right panel shows improvements in the inverse error in the growth
  index, $\gFoM$.  }
\label{fig:FoM_over_FoM0}
\end{figure*}

Our baseline is the combined pre-JDEM BAO+SNIa+WL+Planck case from the FoMSWG
report. The baseline uncertainties in various dark energy parameters, after
marginalization over all nuisance and cosmological parameters, are
$\sigma(w_0) = 0.10$, $\sigma(w_a) = 0.31$, $\sigma(w_p)=0.028$ ($z_p=0.40$),
$\FoM=116$, and $\sigma(\gamma)=0.21$ ($\gFoM=4.8$).  These constraints are
dominated by BAO+Planck which alone yield $\sigma(w_0) = 0.15$, $\sigma(w_a) =
0.44$, $\sigma(w_p)=0.037$, 
$\FoM=61$.  In comparison, WL+Planck and SNIa+Planck yield $\FoM=9.8$ and
0.42, respectively.

We first consider the cluster information alone, with only a %weak
Planck prior adopted from \cite{FoMSWG}.  The constraints in this case,
assuming flat external priors on cluster nuisance parameters, are $\sigma(w_0)
= 0.10$, $\sigma(w_a) = 0.41$, $\sigma(w_p)=0.036$ ($z_p=0.28$), $\FoM=66$,
and $\sigma(\gamma)=0.17$ ($\gFoM=6.0$).  These constraints are comparable
(and complementary) to the BAO+Planck constraints.  Fig. \ref{fig:PC_clus}
shows the first three principal components for the combined (OPT+SZ) cluster
survey combined with Planck.  Two cases are shown: completely unknown and
perfectly known nuisance parameters.  In the first case, the first principal
component peaks at $z\sim 0.3$, which is not surprising given that most
clusters are at $z\lesssim 1$: while this peak sensitivity is at lower
redshifts than that for BAO surveys, it is at slightly larger $z$ than the
peak for SNIa.

As seen in Fig.~\ref{fig:PC_clus}, adding priors to nuisance parameters moves
the cluster PC weights to higher $z$.  This is easy to understand: freedom in
the nuisance parameters has progressively more deleterious effects as redshift
increases, as can be deduced from
Eqs.~(\ref{eqn:mbiasdefsz})-(\ref{eqn:msigdef}).  Thus, priors on the nuisance
parameters restore the ability of the survey to probe higher redshifts, and
push the principal components to higher $z$.

Next, we combine the cluster information with BAO+SNIa+WL+Planck.  The amount
of information that clusters contribute is a strong function of the
systematics assumed, in our case, a function of the {\it external priors} on
the nuisance parameters.  We find that clusters provide very significant
improvement in the figures of merit even with uninformative (flat) priors on
the cluster nuisance parameters.  The new clusters+SNIa+WL+BAO+Planck figure
of merit is 206, which is nearly a factor of two better than
BAO+SNIa+WL+Planck alone.  The pivot error is $\sigma(w_p)=0.022$ ($z_p=0.41$)
which is $\sim 25\%$ better.  Constraints on $w_0$ and $w_a$ improve by about
50\% each, since with clusters $\sigma(w_0)=0.065$ and $\sigma(w_a)=0.214$.
Constraints on growth improve by more than a factor of 2, with
$\sigma(\gamma)=0.099$ ($\gFoM=10$).  The main effect of
  including clusters is to provide significant additional information on the
  growth of density perturbations. Of the complementary techniques we
  consider, only weak lensing probes the growth, but our fiducial cluster
  model gives (slightly) stronger constraints on growth than the pre-JDEM
  combination of BAO+SNIa+WL+Planck.

The left panel of Fig.~\ref{fig:PC_combined} shows the best-determined three
principal components for the fiducial pre-JDEM survey adopted from
\cite{FoMSWG} when clusters are added, i.e. clusters+SNIa+WL+BAO+Planck.  For
the cluster survey we assume OPT+SZ with flat external priors on nuisance
parameters. We see that clusters make the total principal components look more
``cluster-like'' (compare to Fig.~\ref{fig:PC_clus}) since they add a lot of
information to the total. 
For the same reason, clusters move the
weight of the best-determined PC toward lower redshifts. 

The right panel of Fig.~\ref{fig:PC_combined} shows the accuracies
$\sigma(\alpha_i)$ with which the coefficients $\alpha_i$ of the principal
components can be measured.  The contribution of clusters becomes more pronounced
for higher PC number, leading to a nearly constant fractional improvement
in error.  With clusters, the fourth eigencoefficient is about as well
constrained as the second eigencoefficient in the baseline case without
clusters.

If external priors on the nuisance parameters are available, the full power of
cluster constraints is even more evident. To model such priors, we scale all
errors by the same fractional value --- for each nuisance parameter $p_i$, we
let $F_{ii}\rightarrow F_{ii}(1+\alpha)$ where $\alpha$ varies from zero (flat
prior) to infinity (sharp prior).  Consequently, the additional information in
each nuisance parameter is a fixed fraction of the original (unmarginalized)
error in the parameter. While other choices are possible for adding priors, we
settle on this simple prescription to illustrate the effects of external
information on nuisance parameters.

The left panel of Fig. \ref{fig:FoM_over_FoM0} shows the ratio of the figure
of merit that includes all probes, $\FoM$, and the $\FoM$ with clusters left
out (that is, $\FoM/\FoM_0$ where $\FoM_0=116$).  We consider three cluster
survey scenarios: an optical survey, an SZ survey with optical follow-up for
photometric redshift measurements only \footnote{For SPT, optical follow-up is
expected from the DES, the Blanco Cosmology Survey (BCS) and the Magellan
Telescope.}, and the cross-calibrated OPT+SZ survey.  On the x-axis we show
the effective prior on the scatter in mass (the quantity
$(\siglnM^2(z))^{1/2}$ from Eqs.~(\ref{eqn:msigdefsz}) and
(\ref{eqn:msigdef})) --- priors are added to other nuisance parameters as
well, we simply do not show them. The plot shows that the OPT+SZ combination
improves the total $\FoM$ by more than a factor of four if the scatter is
known to high precision.  In the more realistic cases where mass scatter (and
other corresponding parameters) are known to finite accuracy from independent
measurements, we still see improvement by factors of $\sim 3$.

Priors on nuisance parameters contribute to the information content only if
they are substantially stronger than the intrinsic (``self-calibrated'')
uncertainties in these parameters; for the scatter in mass, for example, this
implies the knowledge of $(\siglnM^2(z))^{1/2}$ to better than $O(1)$ as
Fig.~\ref{fig:FoM_over_FoM0} shows.  Comparing the OPT, SZ, and OPT+SZ
cases, we see that as prior information approaches zero (high values of
$\sigma_{\rm prior}$), the cross-calibration provides a lot of extra
information relative to OPT or SZ alone.  

The right panel of Fig. \ref{fig:FoM_over_FoM0} shows the corresponding figure
of merit for the growth index $\gamma$.  Even stronger improvements are now
seen, with the $\gamma$ figure of merit increasing between a factor of two
(flat priors) and ten (infinitely sharp priors). However, we caution that
simulations of modified gravity models need to be done to determine whether
the impact of modified structure growth on the cluster abundance is adequately
captured by the $\gamma$ parameter.  Nevertheless, the right panel of Fig.
\ref{fig:FoM_over_FoM0} indicates that clusters appear to have at least as
much potential to improve the pre-JDEM constraints on the growth history of
the universe as they do for the expansion history (a similar conclusion has
been reached in Ref.~\cite{tang_weller} for a specific modified gravity
model).  We also see that the SZ survey is more useful for improving $\gFoM$
than the DETF $\FoM$; this is because SZ probes higher
  redshifts, which allows for improved constraints on the redshift evolution
  of the growth of structure and hence $\gamma$. 

%%%%%%%%%%%%%%%%%%%%%%%%%%%%%%%%%%%%%%%%%%%%%%%%%%%%%%%%%%%%%%%%%%%%%%%%%%%%%%%%%%%%

\section{Discussion: Implications of the Assumptions}\label{sec:disc}

In this section we discuss the validity of the assumptions we made and the
consequences of varying those assumptions.  We divide our assumptions
into optimistic and pessimistic.  

The assumptions we consider optimistic are:\\[-0.6cm]

\begin{itemize}
\item %O1 -
  The optical mass threshold ($\Mth=10^{13.5}h^{-1}\Msun$); \\[-0.6cm]
\item %O2 -
  The SZ mass threshold ($\Mth=10^{14.2}h^{-1}\Msun$);\\[-0.6cm]
\item %O3 -
  Perfect selection for both SZ and optical cluster finding;\\[-0.6cm]
\item %O4 -
  SPT area (4000 sq. deg.; could be less);\\[-0.6cm]
\item %O5 -
  Known functional form of the scatter in the mass-observable relation (lognormal);\\[-0.6cm]
\item %O6 -
  No mass dependence in the SZ mass-observable scatter (see
  Eq.~(\ref{eqn:msigdefsz})).
\item %O6 -
  Perfect knowledge of photometric redshift errors.
\end{itemize}

The  assumptions that are arguably pessimistic are:\\[-0.6cm]

\begin{itemize}
\item %P1 -
  No other cluster techniques (e.g. X-ray or weak lensing) are available to
  further cross-calibrate cluster counts;\\[-0.6cm]
\item %P2 -
  Large fiducial value of scatter for both optical ($\sigma_0=0.5$) and SZ ($\sigma_0=0.25$);\\[-0.6cm]
\item %P3 -
  Area of DES (4,000 sq.\ deg.; could be as large as
  10,000 sq. degrees);\\[-0.6cm]
\item %P4 -
  Low redshift range of optical cluster-finding ($z<1$);\\[-0.6cm]
\item %P5 -
  Cubic polynomial evolution of redshift scatter for optical and SZ and mass
  evolution of optical scatter (Lima \& Hu, 2005 \cite{lim05} show that cubic
  redshift evolution of the scatter yields near-maximal degradation of
  cosmological parameters);\\[-0.6cm]
\item %P6 -
  Constraints are based on our current knowledge of cluster physics, while the
  field is developing rapidly.
\end{itemize}

The first three optimistic assumptions are the most important.  Since the
mass-function falls rapidly with increasing mass, the lower mass bins contain
most of the clusters, and are therefore most relevant.  Cunha (2009)
\cite{cun09} shows that cross-calibration decreases the sensitivity of the
constraints to the mass threshold somewhat.  Here we have checked that
increasing the optical limit from $\log \Mth = 13.5$ to $13.7$, or the SZ
limit from $14.2$ to $14.5$ degrades the figures of merit by
$10$-$20\%$. Increasing both leads to $30\%$ degradation in the FoMs.
The importance of uncertainty in photometric redshift errors has been studied
extensively by \cite{lim07}.
For the surveys we consider here, it is not unreasonable to assume that large
enough training sets will be available to sufficiently constrain the evolution of the redshift
errors and characterize the survey selection.

Of the pessimistic assumptions, the first one is especially significant: for
example, if X-ray information is available (as expected from surveys
such as eRosita), then X-ray plus optical cross-calibration alone can lead to
excellent dark energy constraints even with the unexpected failure of one or
more of our SZ assumptions.

To test assumptions about the functional form of the scatter, we added another
Gaussian to both optical and SZ mass-observable relations\protect{\footnote{Data and
  simulations (see e.g. Cohn \& White 2009 \cite{coh09}) suggest that the
  double Gaussian is a good representation of projection effects.}}, for a
total of 27 new parameters (43 total); the new parameters describe the
evolution with redshift and mass of the mean and variance of the new Gaussian
(cf. Eqs.~\ref{eqn:mbiasdef} and \ref{eqn:msigdef}), the ratio between the two
Gaussians describing each mass-observable relation, and the correlation
coefficient between optical and SZ (see \cite{cun09}). The figures of merit
degrade by merely 15-20\%.  The small additional degradation is a consequence
of the fact that the new nuisance parameters do not introduce significant new
degeneracies with cosmological parameters.  If we instead add 4 parameters to
characterize the {\it mass dependence} of the SZ scatter and bias, the
degradations are even weaker, being $\lesssim 5\%$.  Intuitively, adding
mass-dependent evolution of the SZ bias and scatter is not as important as the
functional form of the scatter because the SZ probes too narrow a range of
masses for the evolution to be significant.

The arguments and tests outlined in this section show that the assumptions
made in this paper are not overly optimistic, and that the unforeseen
systematic effects would have to be rather capricious in order to lead to
significant further degradations in the cosmological constraints.

%%%%%%%%%%%%%%%%%%%%%%%%%%%%%%%%%%%%%%%%%%%%%%%%%%%%%%%%%%%%%%%%%%%%%%%%%%%%%%%%%%%%
\section{Conclusions}\label{sec:conc}

We have shown that galaxy clusters are a potentially powerful complement to
other probes of dark energy. Assuming optimally combined optical and SZ
cluster surveys based on fiducial DES and SPT expectations and allowing for a
generous set of systematic errors (a total of 16 nuisance parameters), we have
shown that the constraints on the figure of merit expected in 2016 from baryon
acoustic oscillations, type Ia supernovae, weak lensing, and Planck improves
by nearly a factor of two when clusters are added. 
This improvement is achieved without any external prior knowledge on the 
cluster mass-observable nuisance parameters (but also without explicitly allowing 
for errors in the theoretically predicted mass function or cluster
selection).

We have further illustrated the cluster contribution to constraints by
computing, for the first time, the principal components of the equation of
state of dark energy for clusters alone and clusters combined with other
probes. We found that the first cluster principal component peaks at $z\simeq
0.3$, indicating the ``sweet spot'' of cluster sensitivity to dark
energy. This redshift increases slightly if external information on the
cluster nuisance parameters is available. Each eigencoefficient of the
principal component expansion is improved by about a factor of two when
clusters are added, indicating that the improvements extend to well beyond one
or two parameters.

Finally, we have shown that measurements of the growth index of linear
perturbations $\gamma$ (which is a proxy for testing modified gravity) improve
by a factor of several with cluster information. While this particular
calculation depends on assumptions about the modified gravity model, it
broadly illustrates the intrinsic power of clusters to measure growth and
distance separately and to obtain useful constraints on modified gravity
explanations for the accelerating universe.

We conclude that cross-calibrated cluster counts have enough intrinsic
information to significantly improve constraints on dark energy even if the
associated systematics are not precisely known.

%%%%%%%%%%%%%%%%%%%%%%%%%%%%%%%%%%%%%%%%%%%%%%%%%%%%%%%%%%%%%%%%%%%%%%%%%%%%%%%%%%%%

\vspace{-0.8cm}
\acknowledgments 
\vspace{-0.3cm}
CC and DH are supported by the DOE OJI grant under contract
DE-FG02-95ER40899, NSF under contract AST-0807564, and NASA under contract
NNX09AC89G.  DH thanks the Galileo Galilei Institute in Firenze for good
coffee, and we thank an anonymous referee, Gus Evrard and Eduardo Rozo for comments.

\bibliography{pca.clusters}

\end{document}